\documentclass[pra,aps,twocolumn]{revtex4}

\usepackage{epsfig}
\usepackage{amssymb}

\begin{document}

\sloppy

\title{Fresnel Representation of the Wigner Function: An Operational Approach}

\author{
	P. Lougovski$^{1,2}$, 
	E. Solano$^{1,3}$, 
	Z. M. Zhang$^{1,4}$, 
	H. Walther$^{1,2}$, 
	H. Mack$^{5}$, 
	and W. P. Schleich$^{5}$}

\affiliation{
	$^1$Max-Planck-Institut f{\"u}r Quantenoptik, Hans-Kopfermann-Strasse 1, D-85748 Garching, Germany\\
	\hspace*{-1cm}$^2$Sektion Physik, Ludwig-Maximilians-Universit\"at M\"unchen, Am Coulombwall 1, D-85748 Garching, Germany\hspace*{-1cm}\\
	$^3$Secci\'{o}n F\'{\i}sica, Departamento de Ciencias, Pontificia Universidad Cat\'{o}lica del Per\'{u}, Apartado 1761, Lima, Peru\\
	$^4$Department of Physics, Shanghai Jiao Tong University, Shanghai 200030, China\\
	$^5$Abteilung f\"ur Quantenphysik, Universit\"at Ulm, D-89069 Ulm, Germany
}

\begin{abstract}
We present an operational definition of the Wigner function. Our method relies on the Fresnel transform of measured Rabi oscillations and applies to motional states of trapped atoms as well as to field states in cavities. We illustrate this technique using data from recent experiments in ion traps {[}D. M. Meekhof et al., Phys. Rev. Lett. {\bf 76}, 1796 (1996){]} and in cavity QED {[}B. Varcoe et al., Nature {\bf 403}, 743 (2000){]}. The values of the Wigner functions of the underlying states at the origin of phase space are $W_{|0\rangle}(0)=+1.75$ for the vibrational ground state and $W_{|1\rangle}(0)=-1.4$ for the one-photon number state. We generalize this method to wave packets in arbitrary potentials.
\end{abstract}


\maketitle

Currently the Wigner function~\cite{Hillery} enjoys a renaissance in many branches of physics, ranging from quantum optics~\cite{Quantumoptics}, nuclear~\cite{Feldmeier} and solid state physics to quantum chaos~\cite{Berry}. This renewed interest has triggered a search for operational definitions~\cite{Lamb} of the Wigner function, that is, definitions which are based on experimental setups~\cite{Smithey}. In the present paper we propose such a definition  based on the resonant interaction of an atom with a single mode of the electromagnetic field. In contrast to earlier work it is the atomic dynamics that performs the major part of the reconstruction. Our definition is well-suited for the Jaynes-Cummings model~\cite{Schleich} but allows a generalization to other quantum systems.

Many methods to reconstruct the Wigner function~\cite{Schleich} of a cavity field or the motional state of a harmonic oscillator have been proposed~\cite{Wodkiewicz}. In particular, the Wigner function of a cavity field can be expressed in terms of the measured atomic inversion~\cite{Lutterbach}. This operational scheme~\cite{Nogues} lives off the dispersive interaction between the atom and the field. Since this scheme requires long interaction times, an operational definition of the Wigner function based on a resonant interaction is desirable. The method of nonlinear homodyning~\cite{Wilkens} and quantum state endoscopy~\cite{Bardroff} fulfill this need, but require rather complicated reconstruction schemes.

In contrast, the Fresnel representation proposed in the present paper is rather elementary. It expresses the value of the Wigner function at a phase space point $\alpha$ as a weighted time integral of the measured atomic dynamics caused by the state displaced by $\alpha$. The weight function is the Fresnel phase factor. Since this method can be applied to a cavity field as well as a trapped ion~\cite{Leibfried} we use a harmonic oscillator as a model system. However, we show that this approach can easily be generalized to non-harmonic oscillators.

Our definition relies on controlled displacements~\cite{Leibfried,Brune} of the quantum state of interest and the observation of Rabi oscillations of a two-level atom interacting resonantly with this field. We record the probability~\cite{Bodendorf}
\begin{equation}
\label{Pg} P_g(\tau; \alpha) = \frac{1}{2} - \frac{1}{2}
\sum_{n=0}^{\infty} P_n ( \alpha ) \cos{(2\sqrt{ n + 1}\tau )
}
\end{equation}
of finding the atom in the ground state $|g\rangle$ as a function of dimensionless interaction time $\tau$ and complex-valued displacement $\alpha$. Here $P_n(\alpha)\equiv\langle n|\hat{D}(\alpha)\hat{\rho}\,\hat{D}^{\dagger}(\alpha)|n\rangle$ denotes the occupation statistics of the state $\hat{\rho}$ displaced by the displacement operator $\hat{D}(\alpha)$.

\begin{figure}[!t]
\begin{center}
\includegraphics[width=68mm]{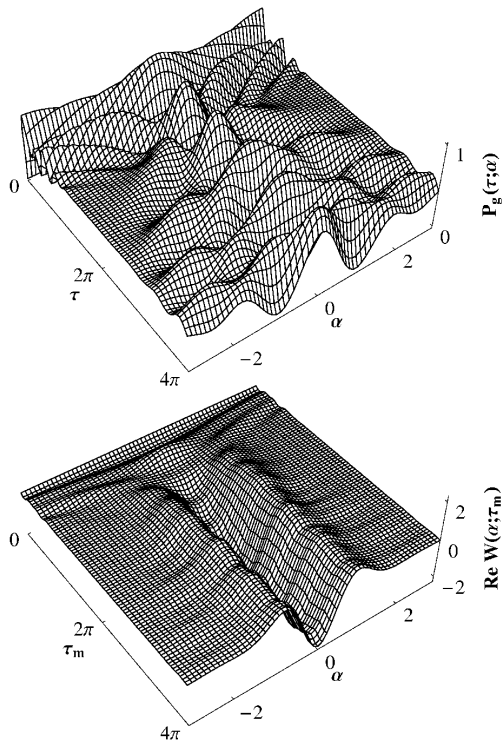}
\caption{\label{fresnelreconstruction}Fresnel
representation of Wigner function illustrated by the first excited
energy eigenstate of a harmonic oscillator. The atomic dynamics
$P_g(\tau;\alpha)$, Eq.~(\ref{Pg}), as a function of interaction 
time $\tau$ and real-valued displacement
$\alpha$ (top) serves as input to the truncated Fresnel representation,
Eq.~(\ref{cutwigner}). For a finite measurement time $\tau_m$ the
so-obtained function $W(\alpha;\tau_m)$ is complex-valued. The
real part of $W(\alpha;\tau_m)$ (bottom) approaches the correct
Wigner function, $W_{|1\rangle}$, already for moderate
measurement times.}
\end{center}
\end{figure}

The Wigner function $W(\alpha)$ of the original
state $\hat\rho$ is determined~\cite{Glauber} by the alternating
sum
\begin{equation}
\label{Wignerfunction} W( \alpha) \equiv
2\sum_{n=0}^{\infty}(-1)^nP_n(-\alpha)
\end{equation}
of the probabilities $P_n$. Hence, we can find the Wigner function
from the atomic dynamics when we measure the probabilities $P_g$
for different interaction times and solve Eq.~(\ref{Pg}) for the
occupation probability $P_n(\alpha)$.

However, there is no need to evaluate the probabilities $P_n$
explicitly~\cite{Santos}. We can obtain the Wigner function
directly from the atomic dynamics $P_g$ without ever calculating
$P_n$, making use of the integral relation
\begin{equation}
\label{aFresnel} \frac2{\pi\sqrt{i}}\int_0^\infty\!\! d\tau
\exp{(i\tau^2/\pi)}\cos(2\sqrt{n}\tau) = (-1)^n.
\end{equation}
When we multiply $P_g-1/2$ from Eq.~(\ref{Pg}) by $2/(\pi\sqrt{i})\exp{\lbrack i \tau^2 / \pi\rbrack}$ and integrate over $\tau$, Eq.~(\ref{aFresnel}) yields
\begin{displaymath}
\frac2{\pi\sqrt{i}} \! \int_0^{\infty} \!\!\!\!\! d\tau e^{i
\tau^2 / \pi} \!\! \left[P_g ( \tau ; -\alpha ) \! - \!
\frac{1}{2} \! \right] \! = \! \frac{1}{2} \sum_{n = 0}^{\infty}
(-1)^n P_n (-\alpha).
\end{displaymath}
We recall the connection, Eq.~(\ref{Wignerfunction}), between $W(\alpha)$ and $P_n(\alpha)$ and find the Wigner function $W(\alpha)\equiv\lim\limits_{\tau_m\to\infty}W(\alpha;\tau_m)$ in terms of the truncated Fresnel transform~\cite{fresneltransform}
\begin{equation}
\label{cutwigner} W (\alpha;\tau_m) \equiv
4\int_0^{\tau_m}\!\!\frac{2\,d\tau}{\pi\sqrt{i}}
e^{i\tau^2/\pi}\left[P_g ( \tau ; -\alpha )
-\frac{1}{2}\right]
\end{equation}
of the Rabi oscillations.

\begin{figure}[!ht]
\begin{center}
\includegraphics[width=65mm]{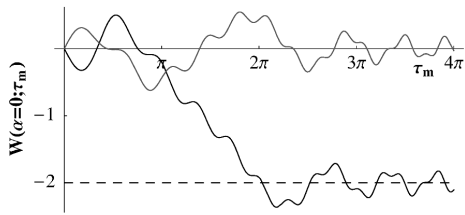}
\caption{\label{cutalphazero}Asymptotic approach
of the truncated Fresnel representation for the first excited state.
Real part (black line) and imaginary part (gray line) of
$W(\alpha=0;\tau_m)$ oscillate around their asymptotic values $-2$
and $0$, respectively, with slowly decreasing amplitudes. The
asymptotic values emerge already after a measurement time of the
order of $2\pi$.}
\end{center}
\end{figure}

Hence, the Wigner function at the phase space point $\alpha$ is
determined by a weighted time integral of the atomic dynamics due to
the initial state displaced by $-\alpha$. The weight function is the Fresnel phase factor $\exp(i\tau^2/\pi)$.

It is instructive to compare the Fresnel representation, Eq.~(\ref{cutwigner}), to the Fourier method used in Ref.~\cite{Leibfried}. The latter relies on the analysis of the Fourier transform of the Rabi oscillations $P_g$ and thus needs the full functional dependence on frequency. In contrast, the Fresnel representation gives the Wigner function as a single integral of $P_g$ without the need to analyze an intermediate function.

The Fresnel representation seems to suffer
from three disadvantages: \textit{(i)} It contains the complete
time evolution of $P_g$, that is, from $\tau=0$ to $\tau=\infty$.
However, any experiment can only record the dynamics for a finite
measurement time $\tau_m$. \textit{(ii)} It can take on complex
values. However, the Wigner function is always real.
\textit{(iii)} It relies on a continuous time evolution. However,
the experiment can only provide a discrete sampling.

We address each of these problems separately and first demonstrate that the phase factor $\exp(i\tau^2/\pi)$ makes the integral insensitive to the long time behavior of $P_g$. For this purpose we show in Fig.~\ref{fresnelreconstruction} the Fresnel reconstruction for the first excited energy eigenstate of a harmonic oscillator. The corresponding Wigner function~\cite{Schleich} $W_{|1\rangle}(\alpha) = -2\left(1-4|\alpha|^2\right)\exp({-2|\alpha|^2})$ is rotationally
symmetric. Therefore, it is sufficient to depict $W_{|1\rangle}$ along the real axis.

The top of Fig.~\ref{fresnelreconstruction} displays the atomic
dynamics $P_g$ as a function of interaction time $\tau$ and
real-valued displacement $\alpha$. Sinusoidal Rabi oscillations
appear along $\alpha=0$. Moreover, for appropriately large
displacements we find collapses and revivals. In the bottom part
we present the real part of $W(\alpha;\tau_m)$. The time axis
corresponds to the upper limit $\tau_m$ of the truncated Fresnel representation,
that is the measurement time. For very short times the function
has no similarity with the correct Wigner function, $W_{| 1
\rangle}$. However, for longer measurement times the curves
approach the exact Wigner function.

\begin{figure}
\begin{center}
\includegraphics[width=85mm]{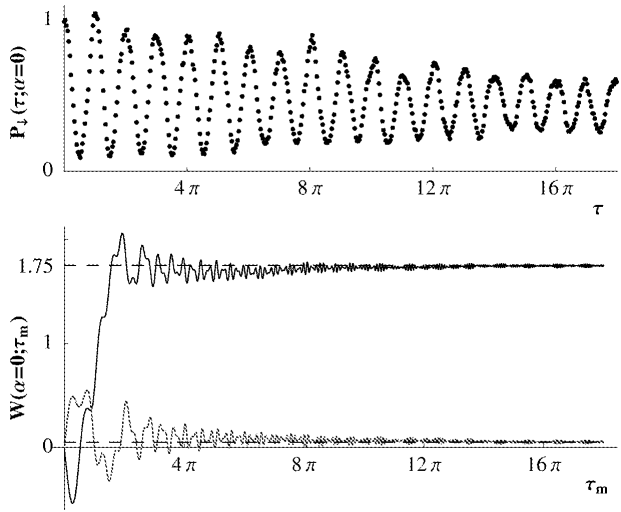}
\caption{\label{fig3}Wigner function at the origin of phase space obtained from experimental Rabi oscillations (top) of a stored ion~\cite{Meekhof} using the truncated Fresnel representation, Eq.~(\ref{cutwigner}): Approach (bottom) of real (black line) and imaginary parts (gray line) of $W(\alpha=0;\tau_m)$ towards their asymptotic values (dashed line) yielding $W_{|0\rangle}(0)=1.75+i0.05$. The ion was prepared initially in the atomic and vibrational ground state.}
\end{center}
\end{figure}

In order to study the asymptotic behavior in more detail we show
in Fig.~\ref{cutalphazero} a cut of the truncated Fresnel
representation along $\alpha=0$. Here we depict both, real and
imaginary part of $W(\alpha;\tau_m)$. We note that after
$\tau_m\approx2\pi$ the real part, indicated by the black line,
begins to oscillate around the correct value
$W_{|1\rangle}(\alpha=0)=-2$ with decreasing amplitude. In a
similar way, the imaginary part, denoted by the gray line,
approaches its asymptotic value zero. Unfortunately, this
convergence is slow. This feature is a consequence of the familiar
Cornu spiral~\cite{Schleich}, also used by R. P. Feynman and J. A.
Wheeler to add up scattered waves --- a precursor of the path
integral~\cite{Feynman-Wheeler}. Despite the slow asymptotics the
knowledge of the Rabi oscillations up to a measurement time of a
least $\tau_m\cong2\pi$ is sufficient to obtain the asymptotic
Wigner function.

The upper part of Fig.~\ref{fig3} shows measured Rabi oscillations of an ion due to its center-of-mass motion~\cite{Meekhof,AntiJC}. Due to the high sampling frequency we can directly interpolate the measured data to evaluate the Wigner function. In the bottom part we display the approaches of the real part (black line) and imaginary part (gray line) of the truncated Fresnel integral towards their asymptotic values (dashed lines). For the ion in the motional ground state with Wigner function~\cite{Schleich} $W_{|0\rangle}(\alpha)=2\,\exp({-2|\alpha|^2})$ we find the asymptotic value $W_{|0\rangle}(0)=+1.75+i\ 0.05$ whereas the ideal value is $W_{|0\rangle}(0)=2$. The presence of an imaginary part in the estimated Wigner function does not represent a fundamental drawback, as long as it comes from a complex integral of realistic data that ideally should produce a real value.

It is quite remarkable that the method works so well considering the fact that in the particular experiment~\cite{Meekhof} the atomic dynamics is governed by the non-linear Jaynes-Cummings model~\cite{Vogel}. In this case the relation between the atomic population and the photon statistics is not of the form of Eq.~(\ref{Pg}). However, for low excitations the deviations from the familiar Jaynes-Cummings model are small as shown in Fig.~1(b) of Ref.~\cite{Meekhof}. Therefore, superpositions of low excitations can be successfully reconstructed.

\begin{figure}
\begin{center}
\includegraphics[width=75mm]{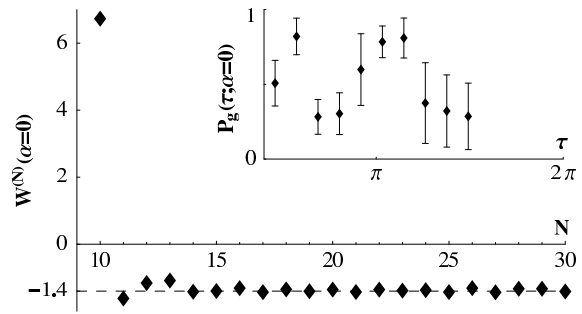}
\caption{\label{fig4}Dependence of the Wigner function $W^{(N)}(\alpha=0)$ at the origin of phase space on the cutoff $N$ of the overdetermined system, Eq.~(\ref{lineq}). The values scatter around $W_{|1\rangle}(0)=-1.4$. The Rabi oscillations~\cite{Varcoe} (inset) result from a cavity field prepared in a one-photon number state.}
\end{center}
\end{figure}

The inset of Fig.~\ref{fig4} shows Rabi oscillations of an atom due to a cavity field~\cite{Varcoe}. Since in this case only a few data points are available we have to employ a discrete version of the Fresnel representation. The data~\cite{Varcoe} consist of a finite set of $M$ interaction times $\tau_j$ and measured probabilities $P_g(\tau_j;-\alpha)$. We now choose kernel coefficients $f_j$ such that
\begin{equation}\label{lineq}
	\sum\limits_{j=1}^M\cos(2\sqrt{n+1}\tau_j)\,f_j=(-1)^{n+1},
\end{equation}
multiply Eq.~(\ref{Pg}) evaluated at discrete times $\tau_j$ by $f_j$, and sum over $j$. With the help of Eq.~(\ref{lineq}) we then arrive at the discrete representation
\begin{equation}\label{wignerdiscrete}
	W(\alpha)=4\sum\limits_{j=1}^Mf_j\left[P_g(\tau_j;-\alpha)-\frac12\right]
\end{equation}
of the Wigner function.

Since the photon number $n$ runs from zero to infinity, Eq.~(\ref{lineq}) describes a system of infinitely many equations for $M$ unknown coefficients $f_j$. When we consider the truncated system of $N>M$ equations we find the solutions $f_j^{(N)}$ of this overdetermined system by minimizing the sum of quadratic deviations. Substitution of the so-calculated coefficients $f_j^{(N)}$ into Eq.~(\ref{wignerdiscrete}) results in the Wigner function $W^{(N)}(\alpha)$.

Figure~\ref{fig4} displays the dependence of $W^{(N)}(\alpha=0)$ on the cutoff $N$. The values scatter around $W_{|1\rangle}(0)=-1.4$ indicated by the dashed line. We recall that the ideal value of the Wigner function corresponding to a one-photon number state is $W_{|1\rangle}(0)=-2$. For $N=M=10$ the method is extremely sensitive to uncertainties in the measured data resulting in a large deviation from -1.4.

The error bars $\delta P_j$ translate into a standard deviation $\delta W=4\left[\sum_j f_j^2\delta P_j^2\right]^{1/2}$ of the Wigner function. According to Eq.~(\ref{lineq}) the coefficients $f_j$ are solely determined by the set of $\tau_j$ and can take almost any values. Obviously large values give rise to a large error in the Wigner function, unless the corresponding uncertainties $\delta P_j$ are small and compensate this effect.

For the experiment of Ref.~\cite{Varcoe} performed for a different purpose and therefore not taking advantage of this feature, we find $\delta W\approx4.3$. However, a future experiment can either choose the interaction times $\tau_j$ such as to minimize the $f_j$ or increase the accuracy of the measurement of $P_g$ at times where $f_j$ is large.

The Fresnel representation, Eq.~(\ref{cutwigner}), of the Wigner
function as well as the discrete version, Eq.~(\ref{wignerdiscrete}), rely on the resonant Jaynes-Cummings model. However, the technique to obtain the Wigner function directly by an
appropriate integral transform of the time dependence of a
measurable quantity is much more general. Indeed, our method even
allows us to obtain the Wigner function of a wave packet $| \psi
(0) \rangle = \sum_n \psi_n | \phi_n \rangle$ consisting of a
superposition of energy eigenstates $| \phi_n \rangle$ with energy
$E_n \equiv \hbar \omega_n$ of a potential $V \equiv V (\vec{r})$.
Here, we do not restrict ourselves to a harmonic oscillator
potential. Our only assumption is that $V$ is symmetric with
respect to the origin. In this case, the representation
Eq.~(\ref{Wignerfunction}) of the Wigner function still holds
true. Moreover, we recall that the autocorrelation function ${\cal
C}(t) \equiv \langle \psi(0) | \psi (t) \rangle = \sum_n
|\psi_n|^2 \exp(-i \omega_n t)$ is measured routinely in wave
packet experiments, e.g.~\cite{wavepackets}.

When we apply the displacement $\hat{D}$ to the initial state
$\rho \equiv | \psi (0) \rangle \langle \psi (0) |$, the
autocorrelation function ${\cal C} (t; \alpha) = \sum_n P_n (\alpha)
\exp(-i \omega_n t)$ is similar to Eq.~(\ref{Pg}) with $P_n(\alpha) \equiv \langle
\phi_n | \hat{D} (\alpha) \hat{\rho} \hat{D}^{\dagger} (\alpha) |
\phi_n \rangle$. We multiply both sides of the equation for ${\cal C}$ by
\begin{eqnarray}
\label{kernel} f(t) \equiv \frac{1}{2 \pi} \int_{-
\infty}^{\infty} d \omega \, e^{i \omega t} \cos \lbrack n(\omega)
\pi \rbrack
\end{eqnarray}
and integrate over $t$. Here $n(\omega)$ is an arbitrary
and continous function such that $n(\omega_n) = n$.

When we make use of the symmetry relations $f(-t) = f^* (t)$ and ${\cal C} (-t)
= {\cal C}^* (t)$, we find the Wigner representation
\begin{eqnarray}
\label{generalWigner} W(\alpha) = 4\,Re \left[ \int_{0}^{\infty}
dt f(t) {\cal C} (t; -\alpha) \right]
\end{eqnarray}
of the initial wave packet in terms of the measured
autocorrelation function ${\cal C}$ and the integral kernel $f$.

Equation~(\ref{generalWigner}) is the generalization of the
Fresnel representation, Eq.~(\ref{cutwigner}), to a  quantum system
with arbitrary discrete spectrum. For wave packets in a harmonic
oscillator or in a box with energy spectra $\omega_n \equiv  2 \pi
n / T_{cl}$ or $\omega_n \equiv 2 \pi n^2 / T_r$, respectively,
the integral kernels from Eq.~(\ref{kernel}) are delta functions
at half of the classical period $T_{cl}$, or half of the revival
time $T_r$. For these potentials, the time evolution itself
creates the function $(-1)^n$, that is the parity operator. For a
spectrum $\omega_n\equiv\sqrt{n+1}\,2\Omega$,
Eq.~(\ref{kernel}) predicts a quadratic Fresnel-like phase.

Complications arise for spectra with degeneracies, such as the hydrogen atom. Nevertheless, we can reconstruct Rydberg wave packets measuring the autocorrelation function at half of the revival time. These and other relevant examples will be discussed elsewhere. They suggest a wide application of integral (Fresnel) transforms in the search for a practical reconstruction of the Wigner function.

We thank M.~Freyberger for fruitful discussions and D.~Wineland
for allowing us to use his data. The work of H.~M. and W.~P.~S. is
partially supported by the Deutsche Forschungsgemeinschaft. H.~W.
and W.~P.~S. acknowledge financial support from the European
Commission through the IHP Research Training Network QUEST and the
IST project QUBITS.


\begin{thebibliography}{99}
\bibitem{AntiJC} For the anti-Jaynes-Cummings model~\cite{Meekhof}, the minus sign
in Eq.~(\ref{Pg}) is replaced by a plus sign and the Fresnel
representation, Eq.~(\ref{cutwigner}) assumes an overall minus sign.
\end{thebibliography}
\end{document}